\documentclass[a4paper]{jpconf}
\usepackage{graphicx}
\usepackage{amsmath}

\newcommand\la         {{\langle}}
\newcommand\ra         {{\rangle}}
\newcommand\rd         {{\mathrm d}}
\newcommand\rD         {\Delta}

\newcommand\sproc      {$s$ process}
\newcommand\rproc      {$r$ process}

\newcommand\adecay     {$\alpha$ decay}
\newcommand\bdecay     {$\beta$ decay}
\newcommand\sdproc     {$s$-process}

\newcommand\addecay    {$\alpha$-decay}
\newcommand\bddecay    {$\beta$-decay}

\newcommand\Ref[1]     {Ref.\,\cite{#1}}

\newcommand\eqn[1]     {Eq.\,(\ref{#1})}
\newcommand\eqns[2]    {Eqs.\,(\ref{#1}) and~(\ref{#2})}

\newcommand\fig[1]     {Fig.\,{\ref{#1}}}

\newcommand\nn         {\nonumber}

\def\beq{\begin{equation}}
\def\eeq{\end{equation}}
\def\bsp#1\esp{\begin{split}#1\end{split}}
\def\bal#1\eal{\begin{align}#1\end{align}}
\def\beeq{\begin{eqnarray}}
\def\eeeq{\end{eqnarray}}

\begin{document}
\title{A unified model for nucleosynthesis of heavy elements in stars}

\author{Mikl\'os Kiss$^1$ and Zolt\'an Tr\'ocs\'anyi$^2$}

\address{$^1$Berze N.J. Gimn\'azium, Kossuth 33, H-3200 Gy\"ongy\"os,
Hungary}
\address{$^2$University of Debrecen and Institute of Nuclear Research of
the Hungarian Academy of Sciences, H-4001 Debrecen P.O.Box 51, Hungary}

\ead{kiss-m@chello.hu, z.trocsanyi@atomki.hu}

\begin{abstract}
We prospose a unified model for the nucleosynthesis of heavy ($A > 57$)
elements in stars. The neutron flux can be set to describe neutron
capture in arbitrary neutron flux.  Our approach solves the coupled
differential equations, that describe the neutron capture and decays of
2696 nuclei, numerically without truncating those to include only
either capture or decay as traditionally assumed in weak neutron
flux (\sproc). As a result the synthesis of heavy nuclei always evolves
along a wide band in the valley of stable nuclei.  The observed
abundances in the Solar system are reproduced reasonably already in the
simplest version of the model. The model predicts that the
nucleosynthesis in weak or modest neutron flux produces elements that
are traditionally assumed to result in the high neutron flux of
supernovae explosions (\rproc).
\end{abstract}

\vspace*{-2em}
\section{Introduction}

The traditional approach for nucleosynthesis in stars that maintain
weak neutron flux is the slow neutron capture process (\sproc) that
occurs along a path in the stability valley of nuclei. After the
pioneering works of Burbidge et al. \cite{Burbidge:1957vc} and of Cameron
\cite{Cameron:1958vx}, this process has been studied extensively in the
literature \cite{Kappeler:1989}. 
The \sdproc\ approach starts with the coupled differential equations that
describe the change (increase and decrease) of the abundance $N_A$ of a
certain nucleus of atomic number $A$ at time $t$ due to neutron
capture and the decrease of the abundance due to \bdecay,
\beq
\frac{\rd N_A}{\rd t}(t) =
  N_n(t) \la \sigma v \ra_{A-1} N_{A-1}(t)
- N_n(t) \la \sigma v \ra_A N_A(t)
- \Gamma^{(\beta)}_{A} N_{A}(t)
\,,
\label{eq:classicdNA}
\eeq
where $N_n(t)$ is the neutron density at time $t$, $\la \sigma v \ra_A$
is the neutron-capture reaction rate per neutron on a nucleus with mass
number $A$ and $\Gamma^{(\beta)} = \ln 2/T^{(\beta)}_{1/2}$ denotes
the \bddecay\ width of the isotope ($T^{(\beta)}_{1/2}$ is the half
life). For the large number of possible nuclei involved in this
process, the standard way of finding an approximate solution is to
assume that either the capture is much faster than the decay, or vice
versa. With this assumption the nucleosynthesis of heavy elements evolves
along a line in the valley of stable nuclei, called the \sdproc\ path.

The reaction rate depends on time through the variation of temperature.
It is usually assumed that the temperature is constant during neutron
irradiation, the typical value being $T = 3.5\cdot 10^8$\,K,
corresponding to $k T = 30$\,keV.  Then one may use
$\la \sigma v \ra_A = \la \sigma_T \ra_A v_T$ where $\la \sigma_T \ra_A$
is the Maxwellian-averaged cross section of neutron capture, that has
the typical value of 100\,mb, and
$v_T = \sqrt{2 kT/m_n} \simeq 2.4\cdot 10^8$\,cm/s.

The classical \sdproc~model can describe the observed abundances of
\sdproc\ heavy elements surprisingly well provided one assumes two
components of exponentially decaying neutron irradiation, the
larger exposure {\em main} component and the smaller exposure {\em
weak} component.  A usual experimental confirmation of the \sdproc\
abundance is the comparison of the model prediction for the neutron
capture cross section times abundance as a function of the mass number
(see e.g.~Fig.~19.\ in \Ref{Kappeler:1989}.) We note that this
comparison involves only those nuclei which belong to the \sdproc~path.

\section{The model}
\label{sec:model}

While the classical \sdproc\ model gives a simple and fairly accurate
description of the observed abundances of heavy elements, it certainly
has some simplifying features that are worth a closer look. One point is
that no matter how low the probability of a certain capture or decay
process, if the sample of nuclei is sufficiently large, such events will
occur in reality. This means that the evolution of nucleosynthesis along
the \sdproc~path can only be a simple approximation and one has to
discuss carefully the conclusions drawn from the model. With present day
computing capacity it is actually not necessary to make the simplifying
assumptions that lead to the \sdproc~path of evolution. Instead, we choose
to solve the full system of differential equations numerically with some
different kind of simplifying assumptions.

If we choose to keep all terms in \eqn{eq:classicdNA}, then we as well
can include more terms that may have some relevance. If we do not force
the evolution onto a path, then we can find increment of a given abundance
$N_A$ by \adecay\ from an already existing nucleus with two more proton
and neutron number as well as decrement by possible \adecay. This
suggests that actually, the more correct abundance to follow is
$N_{Z,N}(t)$, the abundance of the nuclei of a certain element with
atomic number $Z$. Therefore, instead of \eqn{eq:classicdNA}, we choose
to solve the coupled differential equations with one element as
\beq
\bsp
\frac{\rd N_{Z,N}}{\rd t} &=
  N_n(t) \la \sigma v \ra_{A-1} N_{Z,N-1}(t)
+ \Gamma^{(\beta)}_{Z-1,N+1} N_{Z-1,N+1}(t)
+ \Gamma^{(\alpha)}_{Z+2,N+2} N_{Z+2,N+2}(t)
\\ & 
- N_n(t) \la \sigma v \ra_A N_{Z,N}(t)
- \Gamma^{(\beta)}_{Z,N} N_{Z,N}(t)
- \Gamma^{(\alpha)}_{Z,N} N_{Z,N}(t)
\,,
\esp
\label{eq:newdNA}
\eeq
where $\Gamma^{(\alpha)} = \ln 2/T^{(\alpha)}_{1/2}$ denotes
the \addecay\ width of the nuclei.%
This equation reflects the fact that the spectral analysis for observing
a particular nucleus applies to fixed atomic numbers, not mass numbers.

In order to solve the coupled system of equations numerically, we
introduce discrete time steps $\tau$. If the rate of a certain process
is sufficiently large compared to the time step, then the differential
equations can be linearized  
\beeq
&&
N_{Z,N}(t+\tau) =
  N_{Z,N}(t) 
+ p^{(n)}_{Z,N-1} N_{Z,N-1}(t)
+ p^{(\beta)}_{Z-1,N+1} N_{Z-1,N+1}(t)
+ p^{(\alpha)}_{Z+2,N+2} N_{Z+2,N+2}(t)
\nn\\ && \qquad\qquad\qquad\qquad\qquad
- p^{(n)}_{Z,N} N_{Z,N}(t)
- p^{(\beta)}_{Z,N} N_{Z,N}(t)
- p^{(\alpha)}_{Z,N} N_{Z,N}(t)
\,,
\label{eq:newNA}
\eeeq
where $p^{(x)}_{Z,N} = \tau \lambda^{(x)}_{Z,N}$, with $\lambda^{(x)}$
being the capture or decay rate for process $x$. This equation could be
used only if the time step is (much) smaller than any of the capture or
decay inverse rates, so that all probabilities $p^{(x)}_{Z,N}$ are (much)
smaller than one. This condition would make the computer code uselessly
slow because of the often very small decay life times. In order to be
able to choose a sufficiently large time step, we perform the evolution
in two steps, a neutron capture period followed by a decay period in an
alternating manner.

Given a distribution of abundances at a given moment $t$, $N_{Z,N}(t)$,
in the first step we allow only neutron capture with (possibly time
dependent) probability $p^{(n)}_{Z,N}$, which increases each
$N_{Z,N}(t)$ with $p^{(n)}_{Z,N-1} N_{Z,N-1}(t)$ and decreases them
with $p^{(n)}_{Z,N} N_{Z,N}(t)$, so the net change in the capture period
is
\beq
\rD N_{Z,N}^{\rm capt.} = p^{(n)}_{Z,N-1} N_{Z,N-1} - p^{(n)}_{Z,N} N_{Z,N}
\,,
\label{eq:capture}
\eeq
and we update each value of the abundances accordingly.

In the next time step the nuclei present at the end of the previous
capture period are allowed to decay only. This decay will be described
by the usual exponential decay, and the number of decayed nuclei of a
given isotope at the end of the decay period will be
\beq
\rD N_{Z,N}^{\rm dec.}(t+\tau) = N_{Z,N}(t) [1-\exp(-\Gamma_{Z,N}\,\tau)]
\,.
\label{eq:decay}
\eeq
In \eqn{eq:decay} $\Gamma_{Z,N}$ is the total decay width
of an unstable isotope with atomic number $Z$ and neutron number $N$,
the sum of the partial widths for the various possible decays,
$\Gamma_{Z,N} = \sum_i \Gamma_{Z,N}^{(i)}$. If the half life of the
isotope is long as compared to the time step, we can use the
approximation
\beq
\rD N_{Z,N}^{\rm dec.}(t+\tau) \simeq N_{Z,N}(t) \,\Gamma_{Z,N}\,\tau
\,,
\label{eq:decayexpanded}
\eeq
and the combined effect of the two time steps
(\eqns{eq:capture}{eq:decay}) is equivalent to \eqn{eq:newNA}. If
however, $\Gamma_{Z,N} \tau$ is not small the expansion in
\eqn{eq:decayexpanded} cannot be used. We use \eqn{eq:decayexpanded} if
$\Gamma_{Z,N} \tau < -\ln 0.99$, while \eqn{eq:decay} if
$-\ln 0.99 < \Gamma_{Z,N} \tau < -\ln 0.01$ and
$\rD N_{Z,N}^{\rm dec.}(t+\tau) = N_{Z,N}(t)$ if 
$\Gamma_{Z,N}\,\tau > -\ln 0.01$ (i.e.\ all nuclei decay immediately).
The total number of nuclei does not change, only redistributed in this
step because the nuclei disappearing from a certain type will turn into
other types of nuclei in proportion to the partial widths of the
possible decay modes.

We took the relevant neutron capture cross sections for given $Z$ and
$N$ from \Ref{LLNL:30kev}, the nuclear decay rates from
\Ref{Tuli:2005} and in the region around bismuth we used the
nuclear decay data from \Ref{Tokai}. We included the relevant nuclear
data of 2696 nuclei of which 178 are stable.

\section{Predictions}

\begin{figure}[ht]
\begin{minipage}{75mm}
\includegraphics[width=75mm]{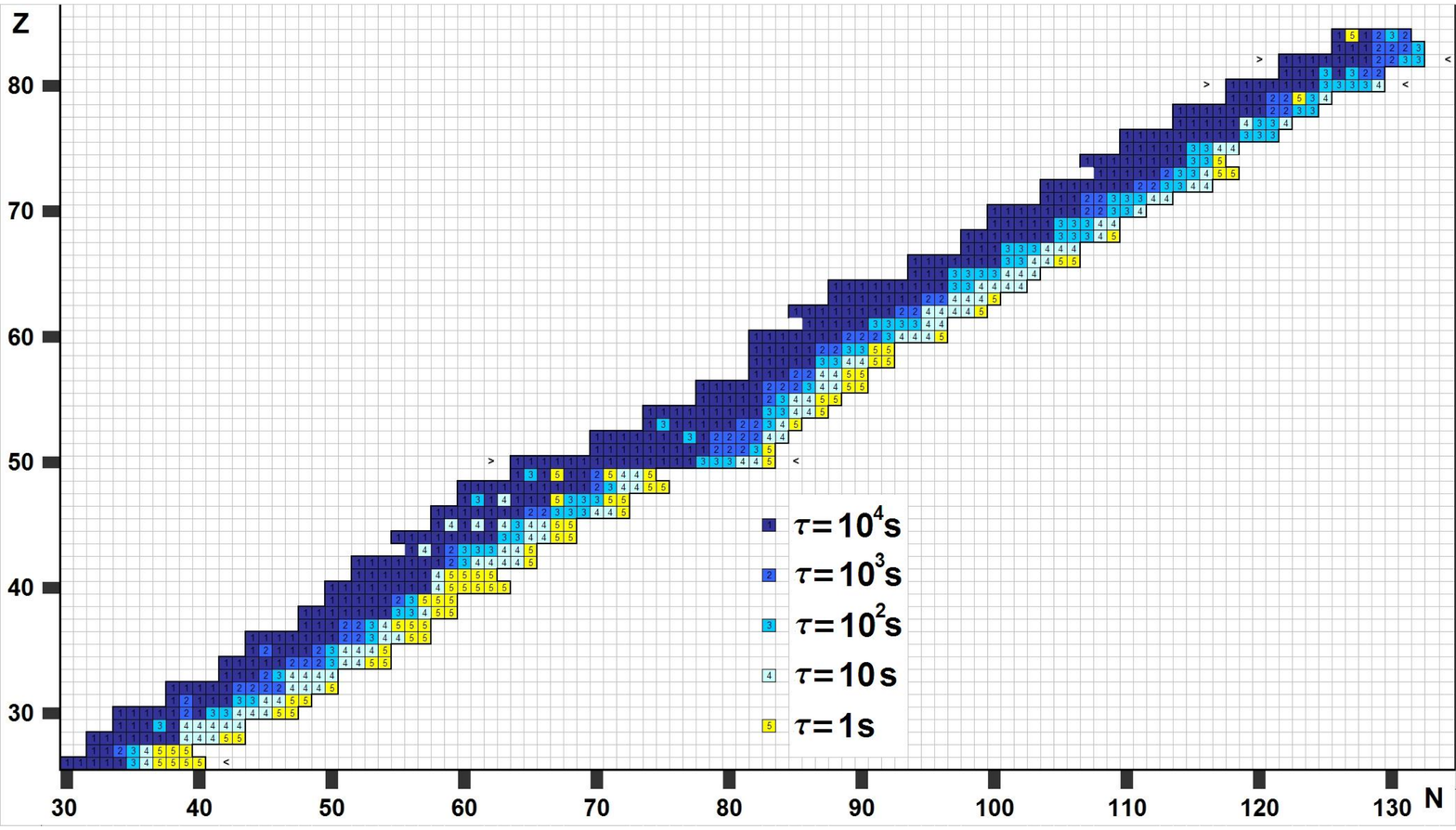}
\caption{\label{fig:maptau}The evolution band obtained with
$\Phi = 2.4\cdot 10^{-11}$\,mb$^{-2}$s$^{-1}$. The numbers in the boxes
denote the cycle in which the given nucleus appeared.}
\end{minipage}
\hfill
\begin{minipage}{75mm}
\includegraphics[width=75mm]{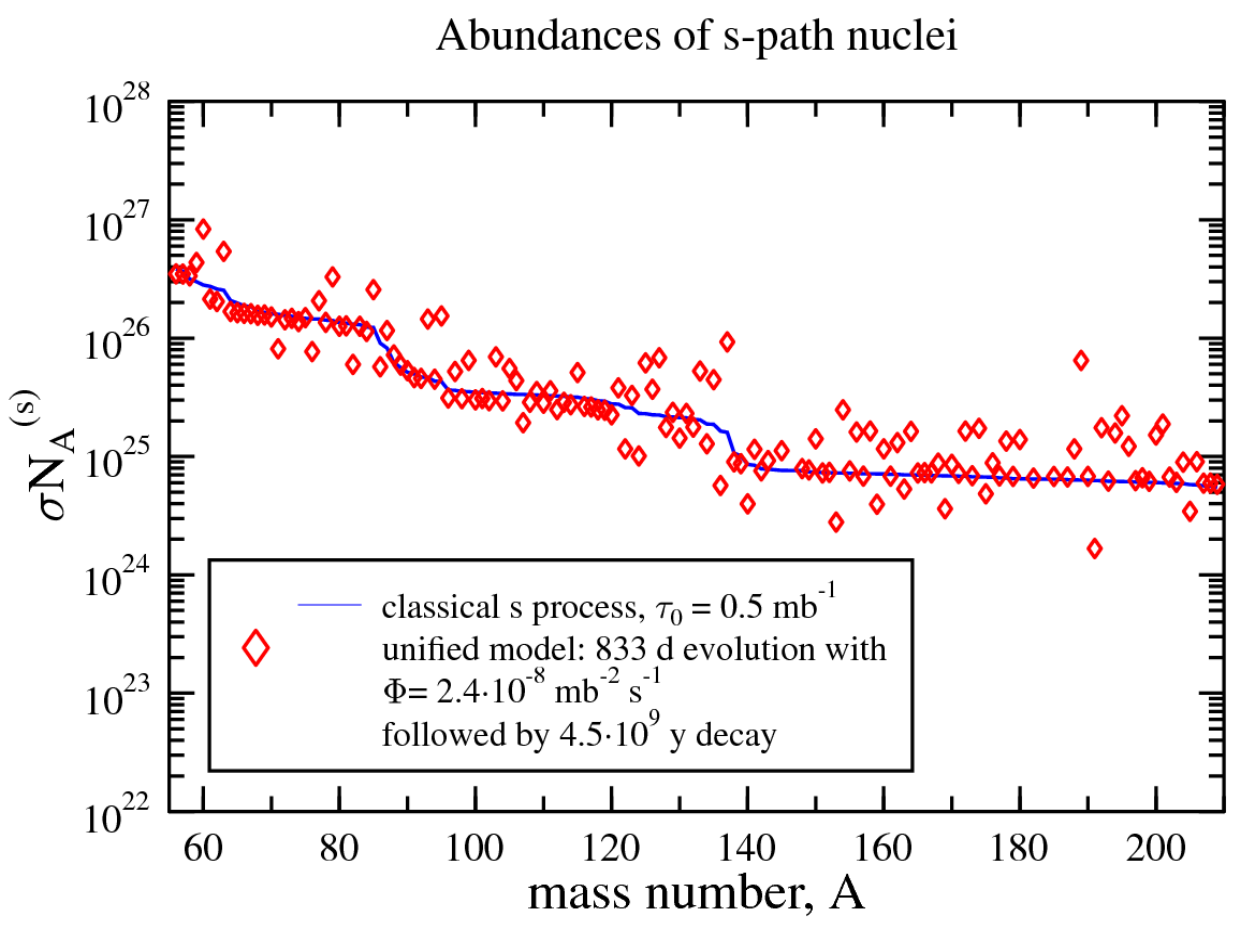}
\caption{\label{fig:sprofile}The abundances of elements obtained with
$\Phi = 2.4\cdot 10^{-8}$\,mb$^{-2}$s$^{-1}$ in 845\,y of evolution and
4.5 billion years of decay compared to the abundances obtained in the
classical \sproc.}
\end{minipage}
\vspace*{-4mm}
\end{figure}

The neutron-capture rate of a nucleus is proportional to its
neutron-capture cross section $\lambda = \Phi \la \sigma_T \ra$, where
$\Phi = N_n v_T$ is the neutron flux. Using the value of the neutron
density typically assumed in the \sproc\ (the case of weak neutron
flux), $N_n = 10^8\,{\rm n\,cm^{-3}}$ and average speed of neutrons,
$v_T = 2.4\cdot 10^8$cm\,s$^{-1}$, the flux is
$\Phi = 2.4 \cdot 10^{-11}$mb$^{-2}$s$^{-1}$. The evolution band
obtained with this parameter is shown on \fig{fig:maptau} after five
cycles of 2.4 million steps with successively decreasing time steps
$\tau_i = 10^{5-i}$\,s ($i = 1,\dots,5$). Thus the total evoulution
time is about 845\,y. The smaller time steps, the wider band. Thus
large time steps approximate the \sproc. The evolution starts with
$3\cdot 10^{45}$ $^{56}$Fe nuclei and ends up producing 1090 nuclei.
The evolution always progresses along a band and increasing the neutron
flux, the band gets wider.

\fig{fig:sprofile} shows the predicted abundances of elements
summed over isobars after two steps: (i) an initial
evolution of 845\,y with $\Phi=2.4\cdot 10^{-8}$\,mb$^{-2}$s$^{-1}$
followed by (ii) a decay of 4.5 billion years.  We compare our
prediction to abundances resulting from the classical \sproc\ (obtained
with the same nuclear input). We see that our model gives identical
results to the classical \sproc.

The nucleosynthesis in our model evolves along a band in the stability
valley until it seemingly stops at the \addecay ing isotopes of
polonium, the $Z = 84$ line. However, with existing neutron flux the
number of heavy nuclei constantly increases. The decay process is
probabilistic, therefore, not all nuclei decay in a given period.  If
the neutron flux is maintained for sufficiently long time, the band
widens below the $Z = 84$ line until it reaches $^{218}$Bi. The time
needed for the appearance of $^{218}$Bi depends on the neutron flux.
For instance, choosing $\Phi = 2.4\cdot 10^{-8}$\,mb$^{-2}$s$^{-1}$
(corresponding to $N_n = 10^{11}\,{\rm n\,cm^{-3}}$), it takes about
one year. The larger the flux, the shorter the required time.  Once
$^{218}$Bi appears, heavier nuclei can easily be produced and appear
quickly up to fermium ($^{260}$Fm).

The rate of the evolution of trans-bismuth elements also correlates
with the neutron flux: the larger $\Phi$, the faster evolution.
After the neutron irradiance ceases, the unstable heavy nuclei
decay quickly. However, some nuclei with long life time, such as $^{235}U$
and $^{238}U$, remain for a long time. The abundances of such nuclei
depend strongly on the neutron density.  For the relatively modest
value of $N_n = 10^{11}\,{\rm n\,cm^{-3}}$ the abundances of uranium isotopes
is negligible as compared to the observable abundances of
elements \cite{Arnett:1996}.
The observed abundances can be obtained with the model with the
parameter value $\Phi \simeq 3\cdot 10^{-5}$\,mb$^{-2}$s$^{-1}$
corresponding to $N_n \simeq 10^{14}\,{\rm n\,cm^{-3}}$
that is still six orders of magnitude smaller than the typical neutron
density in supernovae \cite{Rolfs:1988}.

Another interesting observation is that some elements, that do not
belong to the \sdproc\ path (traditionally called $r$-only nuclei),
also appear in the band with noticible abundances. This means that the
band-like evolution reaches certain $r$-only nuclei even in the case of
weak neutron flux, corresponding to $\Phi=2.4\cdot 10^{-11}$\,mb$^{-2}$s$^{-1}$,
although for such value the abundances of $r$-only nuclei are
negligible.  However, the observed abundances can be obtained with the
still small value of $\Phi \simeq 10^{-9}$\,mb$^{-2}$s$^{-1}$.


We proposed a model of nucleosythesis of heavy elements in
stars. The main features of the model are the following:
(i)~all known neutron-capture and decay data of 2696 nuclei are taken
into account, (ii)~the coupled system of differential equations that
describe the change of abundances of these nuclei is solved numerically.
The model is very simple and does not take into account many effects that
are considered standard in current analyses. Most importantly, the
implementation of time-varying neutron flux is in progress. 

This research was supported by the Hungarian Scientific Research Fund
OTKA K-60432.

\end{document}